\documentclass[aps,prd,twocolumn,superscriptaddress,amsmath,widetext,amssymbepsfig,showpacs]{revtex4}
\usepackage{graphicx}
\usepackage{dcolumn}
\usepackage{bm}
\usepackage{natbib}

\newcommand{\be}{\begin{equation}}
\newcommand{\ee}{\end{equation}}
\newcommand{\bea}{\begin{eqnarray}}
\newcommand{\eea}{\end{eqnarray}}

\newcommand{\LCDM}{$ \Lambda $CDM~}

\newcommand{\mr}[1]{\mathrm{#1}}

\urldef\cosmomc\url{http://www.cosmologist.info/cosmomc}
\urldef\mgcamb\url{http://www.sfu.ca/~gza5/MGCAMB.html}

\begin{document}
\title{New constraints on parametrised modified gravity\\from correlations of the CMB with large scale structure}

\author {Tommaso~Giannantonio}
\affiliation{Argelander--Institut f\"ur Astronomie der Universit\"at Bonn, Auf dem H\"ugel 71, D-53121 Bonn, Germany}

\author {Matteo Martinelli}
\affiliation {Universit\`a degli studi di Roma ``La Sapienza'', P.le Aldo Moro 5, I-00185 Roma, Italy}

\author {Alessandra Silvestri}
\affiliation {Kavli Institute for Astrophysics and Space Research, MIT, Cambridge, MA 02139 USA}

\author {Alessandro Melchiorri}
\affiliation {Universit\`a degli studi di Roma ``La Sapienza'', P.le Aldo Moro 5, I-00185 Roma, Italy}

\begin {abstract}
We study the effects of modified theories of gravity on the cosmic microwave
background (CMB) anisotropies power spectrum, and in particular on its large
scales, where the integrated Sachs--Wolfe (ISW) effect is important. Starting with a general parametrisation, we then specialise to $f(R)$ theories and theories with Yukawa--type interactions between dark matter particles. In these models, the evolution of the metric potentials is altered, and the contribution to the ISW effect can differ significantly from
that in the standard model of cosmology. We proceed to compare these
predictions with observational data for the CMB and the ISW,  performing a
full Monte Carlo Markov chain (MCMC) analysis.  In the case of $f(R)$ theories, the result is an 
upper limit on the lengthscale associated to the extra scalar degree of
freedom characterising these theories. With the addition of data from the
Hubble diagram of Type Ia supernovae, we obtain an upper limit on the lengthscale of the theory of 
$B_0 < 0.4$,  or correspondingly  $\lambda_1 < 1900 \, {\rm{Mpc}}/h $ at $ 95 \%$ c.l.
improving previous CMB constraints. For Yukawa--type models we get a bound on the  coupling $0.75 < \beta_1 < 1.25 $  at the $95\%$ c.l.
We also discuss the implications of
the assumed priors on the estimation of modified gravity parameters, showing that a marginally less conservative choice improves the $f(R)$ constraints to
$\lambda_1 < 1400 \, {\rm{Mpc}}/h $, corresponding to $B_0 < 0.2 $
 at $ 95 \%$ c.l.
\end {abstract}


\maketitle

\section {Introduction} \label {sec:intro}

The observed acceleration of the cosmic expansion poses a challenge for modern
cosmology. Standard general relativity (GR), applied to an expanding universe filled with
radiation and cold dark matter, does not fit the data unless one invokes an
additional component, either a cosmological constant, $\Lambda$, or a
dynamical {\it dark energy} field \cite{Caldwell:2009ix}. The former case corresponds to the well
known standard model of cosmology, or $\Lambda$CDM. An alternative approach to
the phenomenon of cosmic acceleration consists of modifying the laws of
gravity on large scales, in order to allow for self-accelerating
solutions. Well-known examples of such theories are the $f(R)$ models~\cite{Starobinsky:1980te, Capozziello:2003tk, Carroll:2003wy,
  Starobinsky:2007hu, Nojiri:2008nk},  or more general scalar--tensor theories~\cite{EspositoFarese:2000ij, Bartolo:1999sq, Elizalde:2004mq, Chiba:2003ir},
the Dvali-Gabadadze-Porrati (DGP) model~\cite{Dvali:2000hr, Deffayet:2000uy}, and its further extensions such as Degravitation~\cite{Dvali:2007kt}.

Such  theories need to fit the observed data as well as or better than the \LCDM baseline model. This requirement is twofold: at the background level, the predicted expansion history has to be coherent with distance measurements such as those from Supernovae (SNe), the position of the CMB peaks, and baryon acoustic oscillations (BAO); on the other hand, the modified predictions for structure formation need to pass the test of comparison with the observed large scale structure (LSS) of the Universe.
It has been shown for several cases --- for instance for $f(R)$~\cite{Song:2006ej,Bean:2006up,Pogosian:2007sw,Tsujikawa:2007xu,Zhao:2008bn} and DGP~\cite {Lue:2003ky,Koyama:2005kd, Song:2006jk, Song:2007wd, Cardoso:2007xc, Giannantonio:2008qr} --- that the structure formation test can dramatically reduce the parameter space left unconstrained from the background test, and it is therefore instrumental to the quest for a truly viable model.

Structure formation in the Universe is driven by gravity through its potentials, which at the perturbative level in Newtonian gauge are described by two fields $\Psi (x_{\mu})$ and $\Phi (x_{\mu})$, defined as the perturbations to the time-time and space-space parts of the metric respectively. A powerful test of modified gravity is offered by measuring the past evolution of these potentials to compare it with the prediction of a given theory.
In the standard \LCDM model the potentials are expected to be identical, to remain constant during matter domination and to decay at late times, as a consequence of the ongoing transition to the dark energy phase. This is not necessarily the case for other theories.
Metric perturbations 
are not viable
astronomical observables, but we may reconstruct them from observational data which are directly dependent on them~\cite{Jain:2007yk,Song:2008xd}. The most useful  data for this purpose are galaxy and cluster counts, weak lensing and the CMB.

 The CMB is an almost perfectly isotropic black body radiation that has been generated at the epoch of hydrogen recombination and has undergone free streaming since then. Nonetheless, this free streaming may be altered if the CMB photons encounter potential wells which evolve in time. In this case, a non-zero net energy is gained (or lost) by the photons if the potential is becoming shallower (or deeper): this phenomenon, known as the integrated Sachs-Wolfe (ISW) effect~\cite{Sachs:1967er}, leaves an  imprint on the larger scales of the CMB spectrum.
In the standard \LCDM theory, this effect is expected to be generated at late times as a consequence of the potential decay when the background starts accelerating. In models of modified gravity, such as $f(R)$ theories, the magnitude of this late effect can be altered, as well as potentials can evolve during matter domination, therefore generating an ISW effect at earlier redshifts~\cite{Song:2006ej,Pogosian:2007sw}.

The ISW effect can be measured by cross-correlating the CMB with tracers of
the large scale structure (LSS)~\cite{Crittenden:1995ak}, and it has been detected using several
different data sets and was used to constrain cosmology~\cite{Afshordi:2003xu, Cabre:2006qm, Fosalba:2003iy,
  Fosalba:2003ge, Padmanabhan:2004fy, Scranton:2003in, Boughn:2003yz,Gaztanaga:2004sk,Corasaniti:2005pq,
  Giannantonio:2006du, Nolta:2003uy, Ho:2008bz, Xia:2009dr}. The strongest (4.5 $\sigma$) detection to date~\cite{Giannantonio:2008zi} has been obtained by combining multiple data sets at different redshifts, thus exploring the evolution of the potentials in time, and is fully consistent with the \LCDM picture.

Here we focus on $f(R)$ theories and models with a Yukawa--type dark matter interaction, and use the ISW data by~\cite{Giannantonio:2008zi}, in conjunction with the CMB, to test the general parametrisation of modified gravity by exploring the parameter space with a Monte Carlo Markov chain (MCMC) technique.

The plan of the paper is as follows. In Section \ref {sec:theories} first we introduce the parametrisation
used to describe departures from GR, and then specialise to the case of  $f(R)$ and Yukawa--type theories. Then we
review the method of the analysis and the data used in Section \ref{sec:analysis}. In Section \ref {sec:constraints} we give details of our results and in Section \ref{sec:priors} we discuss the effects of the assumed priors on the parameter estimation, before concluding in Section \ref {sec:concl}.

\section {Parametrised Modified Gravity} \label {sec:theories}
Here we describe the formalism we use to parametrise departures from general relativity.

\subsection {Background expansion}
In our analysis we fix the background to that of the
\LCDM model of cosmology. The reasons for this choice are multiple; \LCDM is currently the best 
fit to available data and popular models of modified gravity, e.g. $f(R)$, closely mimic \LCDM at the background level 
with differences which are typically smaller than 
the precision achievable with geometric tests~\cite{Hu:2007nk}. The most significant departures happen at the level of 
growth of structure and, by restricting ourselves to \LCDM backgrounds, we can isolate them. 

\subsection {Structure formation}
In models of modified gravity we expect departures from the standard growth of structure,
even when the expansion history matches exactly the \LCDM one. In general, the rate of clustering of dark matter, as well as the evolution
of the metric potentials, is changed and can be scale-dependent. Moreover, typically there might be an effective anisotropic stress introduced by 
the modifications and the two potentials, $\Phi$ and $\Psi$, are not necessarily equal, as it was the case for \LCDM~\cite{Lue:2003ky,Song:2006ej,Koyama:2005kd, Song:2006jk,Pogosian:2007sw,Tsujikawa:2007xu,Song:2007wd,Daniel:2008et}.
Here we focus  on the effect of the modified evolution of the potential, $\Phi+\Psi$, on the CMB. In order to evolve the potentials we employ the MGCAMB code developed in~\cite{Zhao:2008bn} (and publicly available at \mgcamb) to evaluate the growth of perturbations in models of modified gravity. In this code, the energy-momentum equations remain the standard ones, and the modifications to the Poisson and anisotropy equations are encoded in two functions $\mu(a,k)$ and $\gamma(a,k)$ defined by
 \bea\label{mu}
 &&k^2\Psi=-\frac{a^2}{2M_P^2}\mu(a,k)\rho\Delta \ ,\\
  \label{gamma}
&&\frac{\Phi}{\Psi}=\gamma(a,k) \,,
 \eea
 where $\rho\Delta\equiv\rho\delta+3\frac{aH}{k}(\rho+P)v$ is the comoving density perturbation. 
 
 We will consider theories in which the modifications introduce an effective scalar degree of freedom (d.o.f.), and therefore a characteristic lengthscale. A typical action for such theories would be
\begin{eqnarray}\label{action_gen}
S &=& \int  d^4 x \sqrt {-g} \left[ \frac{M_P^2}{2}R-\frac{1}{2}g^{\mu\nu}\left(\nabla_{\mu}\phi\right)\left(\nabla_{\nu}\phi\right)-V(\phi)\right]\nonumber\\
&&+S_i\left(\chi_i,e^{-\alpha_i(\phi)/M_P}g_{\mu\nu}\right)
\end{eqnarray}
 where $\phi$ represents the scalar d.o.f., $\chi_i$ is the $i^{\textrm{th}}$ matter field and $\alpha_i(\phi)$ is the coupling of $\phi$ to $\chi_i$. We will limit ourselves to cases in which the coupling is a linear function of the scalar field, i.e. $\alpha_i(\phi)\propto\phi$.

For the theories described by the action in Eq.~(\ref{action_gen}), the functions $\mu$ and $\gamma$ can be well represented by the following 
parametrisation introduced by~\cite{Bertschinger:2008zb} (and used in~\cite{Zhao:2008bn})
\bea\label{par_mu_old}
&&\mu(a,k)=\frac{1+\beta_1\lambda_1^2\,k^2a^s}{1+\lambda_1^2\,k^2a^s}\,,\\
\label{par_gamma}
&&\gamma(a,k)=\frac{1+\beta_2\lambda_2^2\,k^2a^s}{1+\lambda_2^2\,k^2a^s}\,,
\eea
where the parameters $\beta_i$ can be thought of as dimensionless couplings, $\lambda_i$ as dimensionful
lengthscales  and $s$ is determined by the time evolution of the characteristic lengthscale of the theory, i.e. the mass of the scalar d.o.f. As shown in~\cite{Zhao:2008bn}, in the case of scalar-tensor theories the parameters 
$\{\beta_i,\lambda_i^2\}$ are related in the following way
\be\label{ST_relation}
\beta_1=\frac{\lambda_2^2}{\lambda_1^2}=2-\beta_2\frac{\lambda_2^2}{\lambda_1^2}
\ee
and $1\lesssim s\lesssim4$.

\subsubsection{$f(R)$ theories}

These theories are a subclass of the models described by the action of Eq.~(\ref{action_gen}), corresponding to the case of a universal fixed coupling $\alpha_i=\sqrt{2/3}\,\phi$, i.e. $\beta_1=4/3$~\cite{Magnano:1993bd}. Moreover, $f(R)$ models that closely mimic \LCDM 
correspond to $s\sim 4$~\cite{Zhao:2008bn}. Therefore the number of free parameters in Eqs.~(\ref{par_mu_old}) and~(\ref{par_gamma})  can be reduced to one, e.g. the lengthscale $\lambda_1$.

 The parametrisation in Eq.~(\ref{par_mu_old})  effectively neglects a factor representing the rescaling of the Newton's constant already at the background level due to the modifications (e.g. $(1+f_R)^{-1}$ in $f(R)$ theories). Such a factor is very close to unity in models that satisfy local tests of gravity~\cite{Hu:2007nk} and, as such, it can be neglected. However, when studying the $f(R)$ case, we need to include it to get a more precise estimate of the ISW effect for the MCMC analysis; therefore we use the following extension of Eq.~(\ref{par_mu_old})
\be\label{par_mu}
\mu(a,k)=\frac{1}{1-1.4 \cdot 10^{-8}|\lambda_1|^2a^3}\frac{1+\frac{4}{3}\lambda_1^2\,k^2a^4}{1+\lambda_1^2\,k^2a^4}\,,
\ee
\noindent where the difference with Eq.~(\ref{par_mu_old}) is the multiplicative factor parametrising $(1+f_R)^{-1}$ which can be expressed in terms of $\lambda_1$~\cite{Hu:2007nk,Oyaizu:2008tb}.
Even with this extended parametrisation,  we are left with a single free parameter, the lengthscale $\lambda_1$.
This parameter can be easily related to the value of the mass scale of the scalar d.o.f. introduced by the addition of the $f(R)$ term to the Einstein-Hilbert action. This scalar d.o.f. is represented by the function $f_R\equiv df/dR$, also dubbed the {\it scalaron}, and $\lambda_1$ corresponds to its mass scale today, i.e. $\lambda_1=1/m_{f_R}^0$. In~\cite{Song:2006ej, Song:2007da} this family of $f(R)$ models was labelled by the
parameter $B_0$  which is related to $\lambda_1$ via $\lambda_1^2=B_0\,c^2/(2H_0^2)$. We will present our results mainly in terms of $B_0$ to facilitate the comparison.

\subsubsection{Yukawa--type dark matter interaction}
Scalar-tensor theories are an example of models with interaction between dark energy and dark matter. The parametrisation in Eqs.~(\ref{par_mu})~and~(\ref{par_gamma}) is well suited also for models where dark matter (DM) particles self-interact \emph{\`a la} Yukawa.  Such models can be described by the presence of an additional mediator, represented by a scalar d.o.f., like the one in the action of Eq.~(\ref{action_gen}), with a time-evolving mass and a coupling to DM particles which is now a free parameter of the model. In this case the only coupling of interest in Eq.~(\ref{action_gen}) would be the one to DM, $\alpha_{\textrm{dm}}\neq 0$. The coupling of this scalar d.o.f. to other matter fields, such as baryons, does not need to be the same of the one to DM; in particular, given the stringent constraints on such a coupling, it can be considered negligible.

 In our parametrisation, the additional mediating force is characterised by the coupling $\beta_1$ (related to $\alpha_{\textrm{dm}}$), the lengthscale $\lambda_1$ and its time evolution $s$. The remaining  parameters, $\beta_2$ and $\lambda_2$, can be related to $\beta_1$ and $\lambda_1$ as in the scalar-tensor case of Eq.~(\ref{ST_relation}). In this case there is no need of any additional factor in the expression for $\mu$, since it takes already into full account the effect of a Yukawa coupling on the clustering of dark matter, (in other words the Yukawa interaction alters the potential produced by a dark matter particle, but does not alter the background Newton's constant). Therefore, when analyzing models with Yukawa--type dark matter interaction we will use Eqs.~(\ref{par_mu})~and~(\ref{par_gamma}) with free parameters $\{\beta_1,\lambda_1,s\}$. 
\\

Armed with the expressions for $\mu$ and $\gamma$ to feed into MGCAMB, we can proceed to evaluate our observables.

\begin {figure}[ht]
\vspace{-1cm}
\includegraphics[width=\linewidth, angle=0]{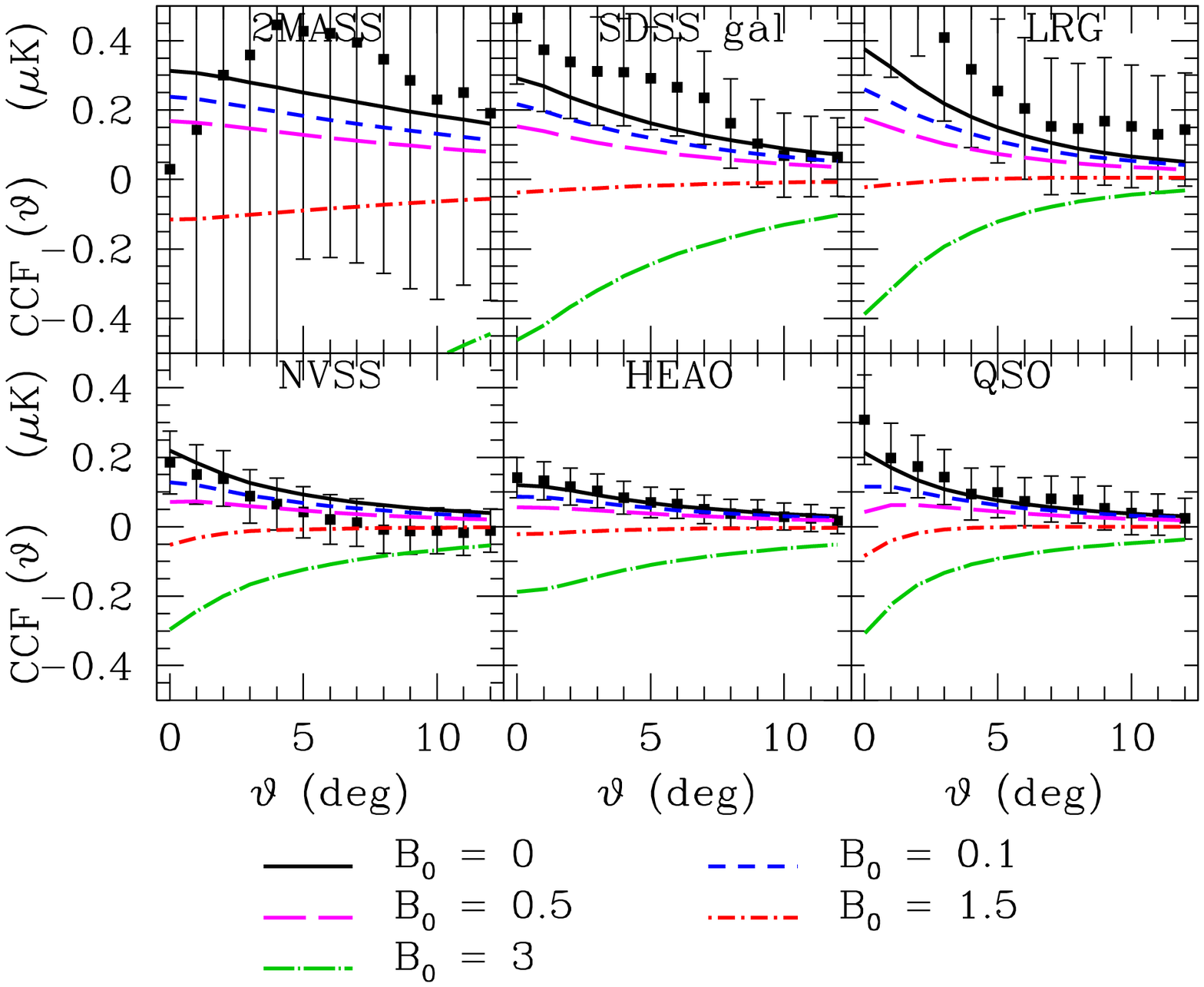}
\includegraphics[width=\linewidth, angle=0]{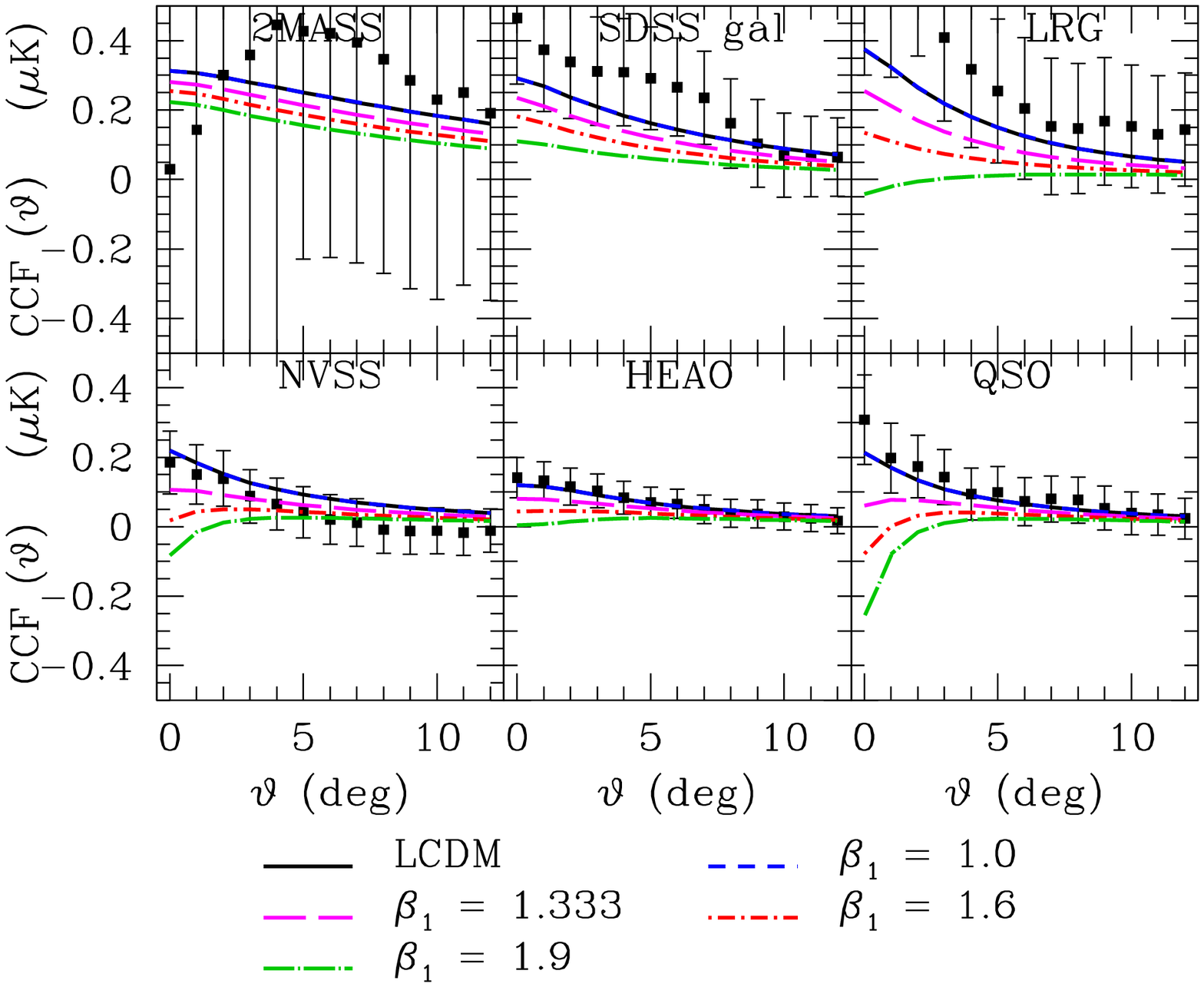}
\caption{Top: theoretical predictions for a family of $f(R)$ theories compared with
  our ISW data~\cite{Giannantonio:2008zi} measuring the angular CCF between
  the CMB and six galaxy catalogues. The model with $B_0 = 0 $ is equivalent to
   \LCDM, while increasing departures from GR produce negative cross-correlations. Bottom: the same for a family of Yukawa--like theories with fixed $B_0=2$. In this case non-unitary coupling generate a redshift evolution of the signal.}
\label{fig:ccf}
\end{figure}

\begin{table*}
\begin{tabular}{|l|l|c| c |}
\hline
Parameter & Explanation & \multicolumn {2}{c|}{range (min, max)}  \\
\hline
 & \multicolumn{1}{c|}{Primary parameters} & $f(R)$ & Yukawa--type \\
\hline
$\omega_b$ & physical baryon density; $\omega_b = h^2\Omega_b$ & \multicolumn {2}{c|}{$(0.005,
0.100)$} \\
$\omega_c$ & physical CDM density; $\omega_c = h^2\Omega_c$ &  \multicolumn {2}{c|}{$(0.01, 0.99)$}  \\
$\vartheta_*$ & sound horizon angle; $\vartheta_* = 100 \cdot r_s(z_\ast)/D_A(z_\ast)$ &
\multicolumn {2}{c|}{$(0.5, 10.0)$ }  \\
 $\tau$ & optical depth to reionisation & \multicolumn {2}{c|}{$(0.01, 0.80)$ } \\
$\Omega_K$ & curvature density; $\Omega_K = 1-\Omega_{\mr{tot}}$ & \multicolumn {2}{c|}{0} \\
$\ln (10^{10} A_s^2)$ & $A_s$ is the scalar primordial amplitude at $k_{\mr{pivot}} = 0.05\,$Mpc$^{-1}$ &  \multicolumn {2}{c|}{$(2.7, 4.0)$} \\
$A_{SZ}$ & amplitude of the SZ template for WMAP and ACBAR & \multicolumn {2}{c|}{$(0, 2)$ } \\
$n_s$ & spectral index of primordial perturbations; $n_s - 1 = d\ln P/d\ln 
k$ &  \multicolumn {2}{c|}{$(0.5, 1.5)$} \\
\hline  
$ B_0   $        &  present lengthscale of the theory (in units of the horizon scale)  &  $(0, 6)$ & $(0, 6)$ \\
$\beta_1$         &   coupling              & $ 4/3 $ & $(0.001, 2)$ \\
$s$               &   time evolution of the scalaron mass             & $ 4 $  & $(1, 4)$ \\
\hline
 & \multicolumn{1}{c|}{Derived parameters} & & \\
\hline
$H_0$ & Hubble parameter [km/s/Mpc]; calculated from $\omega_b$, $\omega_c$,
$\vartheta$, and $\Omega_K$ & \multicolumn {2}{c|}{tophat $(40, 100)$ } \\
$h$ & $h=H_0/(100\,\mbox{km/s/Mpc})$ & \multicolumn {2}{c|}{$(0.40, 1.00)$ } \\
$\Omega_m$ & matter density parameter; $\Omega_m = (\omega_b + \omega_c)/h^2$
& \multicolumn {2}{c|}{} \\
$\Omega_\Lambda$ & vacuum energy density parameter; $\Omega_\Lambda = 1 -
\Omega_K - \Omega_m$ & \multicolumn {2}{c|}{ } \\
\hline
$\lambda_1^2$         & lengthscale of the theory;  $ \lambda_1^2  =   B_0 c^2 / (2 H_0^2)  $   &  $(0, 5 \cdot 10^7)$ & $(0, 5 \cdot 10^7)$ \\
$\lambda_2^2 $        &  second lengthscale of the theory;  $ \lambda_2^2 = \beta_1 \lambda_1^2 $ &  $(0, 7 \cdot 10^7 )$ & $(0, 10^8)$\\
$\beta_2$      & anisotropy parameter;  $2 / \beta_1 - 1$                               & $ 1 / 2
$ & $(0, 2000)$ \\
\hline
\end{tabular}
\caption{Our primary MCMC sampling parameters and some important derived
  parameters. The modified gravity parameters are separated by a horizontal
  line.
 \label{tab:parameters}}
\end{table*}

\section {Analysis} \label{sec:analysis}

\subsection{Method}

We run Monte Carlo Markov chains using the standard Cosmomc package~\cite{Lewis:2002ah},
which is based on the CAMB CMB code~\cite{Lewis:1999bs} and is publicly available at
\cosmomc.  As it is
well known, this technique consists of using a Metropolis-Hastings algorithm
to efficiently sample the multi-dimensional parameter space in which our model
lives. The algorithm is based on multiple chains, which are started from
different random initial points, and then evolved based on a prior probability
distribution of the parameters, assumed flat in the chosen range. Each new step will be accepted if its
likelihood improves the previous likelihood, weighted with some probability.
The final posterior probability distribution is then directly obtained from
the density distribution in the parameter space.

To study our modified gravity (MG)
 model we run five MCMC chains.
We use the parameters described in Table \ref {tab:parameters}: this is based onto
the usual set of \LCDM parameters (note that the angle to the last scattering
surface $\vartheta_{\ast}$ is used instead of the Hubble parameter $H_0$
since this reduces the degeneracies with the other parameters). To account for
$f(R)$ theories we have to add one extra parameter: 
here we chose to use to use $B_0$ as a primary parameter.
 The other
 MG parameters are then fixed:
$\lambda_1^2 = c^2 B_0  / (2 H_0^2)$,
 $ \lambda_2^2 =  \beta_1  \lambda_1^2$,
$\beta_1 = 4/3$, $\beta_2 = 1/2$ and $s
= 4$.
 In theories with Yukawa--type dark matter interaction we have two additional free parameters, i.e. the coupling $\beta_1$ and $s$ as described above. 

We shall discuss the effects of the parameter choice on the priors and on the results in more details in Section \ref{sec:priors}.

\subsection {Data} \label {sec:data}

\subsubsection{CMB}
 For the CMB we use the publicly released WMAP 5 years data~\cite{Nolta:2008ih} for the temperature
and polarisation (TT, TE and
EE) power spectra of the perturbations.
In addition, we used flat top hat priors on the age of the Universe, $ t_0 \in
[10, 20] \mr{Gyrs}$ and on the Hubble parameter today $H_0 \in [40, 100] \mr
{km /s / Mpc} $. 
We include CMB lensing in the analysis.

\subsubsection{ISW}
 To break the degeneracy which remains at the background level between GR and
MG, we use the ISW data by~\cite{Giannantonio:2008zi}.
These data were obtained by cross-correlating the WMAP maps of the CMB with
six galaxy data sets in different bands (2MASS, SDSS main galaxies,
LRGs and QSOs, NVSS, HEAO). The data span different redshift ranges from $\bar
z = 0.1 $ to $\bar z = 1.5$, thus allowing us to study the evolution of
gravity, through the history of the decay of its potentials, in a tomographic way.

There are $78$ actual data points, consisting of the angular cross-correlation
functions (CCFs) in real space, binned at $12$ angles between $0$ and $12$ deg for
each of the catalogues. The full covariance matrix is highly non-diagonal both
between the different angular bins and between the catalogues, due to overlaps
in redshift and sky coverage.
For each MC model, the likelihood contribution due to the ISW data is
calculated as follows: first the theoretical matter and matter-temperature
power spectra $C_l^{gg}, C_l^{Tg}$ are calculated through a full Boltzmann
integration inside CAMB, then a Legendre transformation yields the CCFs and
the matter auto-correlation functions (ACFs) at the
relevant scales. Finally, the bias parameters, assumed constant for each
catalogue, are recalculated for each model by forcing the ACFs
 to match the observations.

We show in Fig. \ref {fig:ccf} the CCF data points, overlapped with the
theoretical predictions for a family of $f(R)$ and Yukawa--like theories. We can see that in the first case a departure from GR produces an unobserved negative signal, while in the second a coupling different from unity causes an equally unobserved redshift evolution of the effect.

\subsubsection{Supernovae}
 To further constrain the background expansion history of the
Universe, we study the effect of including the constraints from the Hubble
diagram of distant Type Ia Supernovae (SNe).  In particular we use the Union SN
compilation by~\cite{Kowalski:2008ez}, which consists of  414 SNe drawn from
13 independent data sets plus 8 newly discovered SNe at low redshift, all reanalysed in a consistent way.

\begin {figure*}[htb]
\includegraphics[width=.33\linewidth, angle=0]{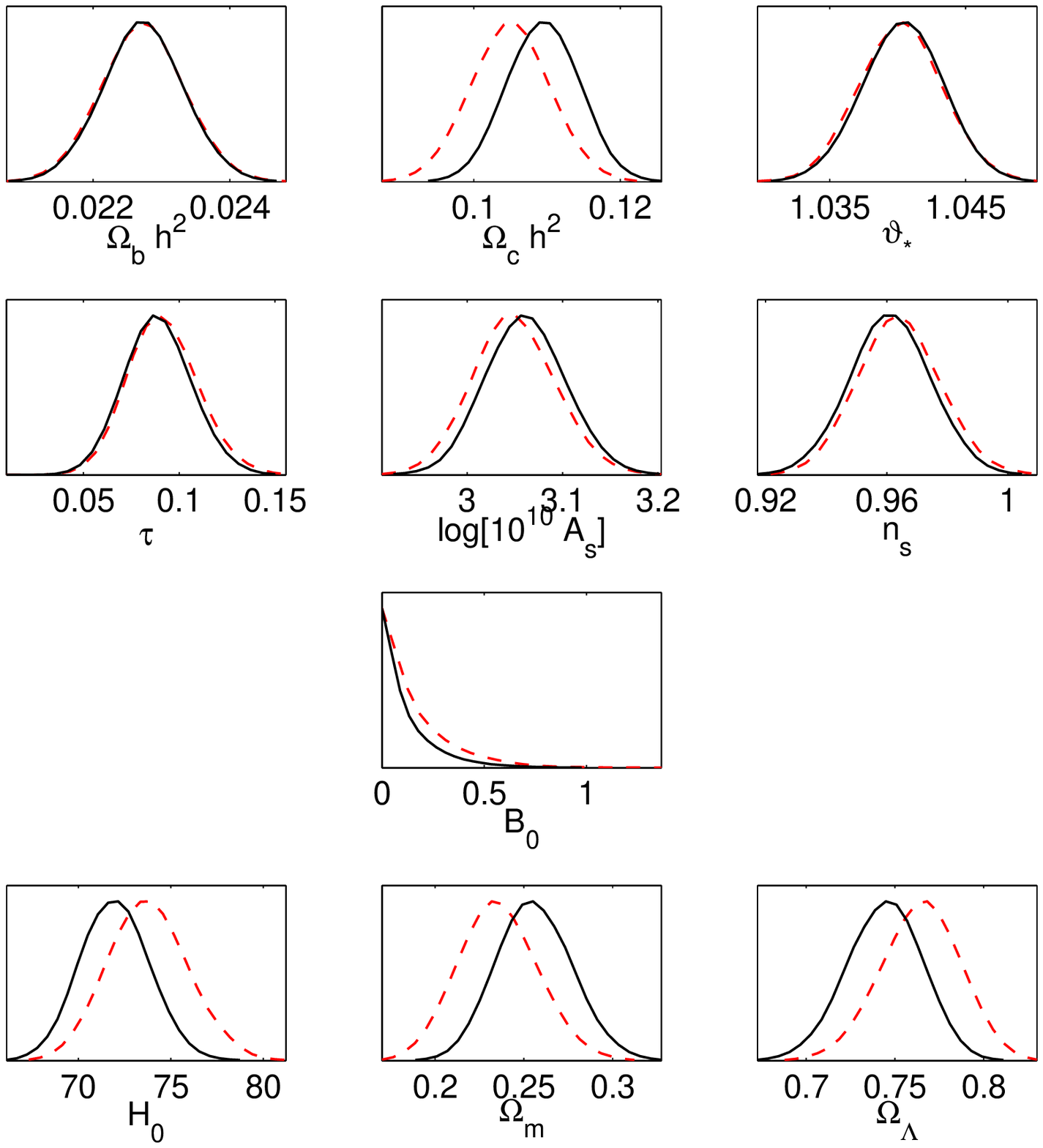}
\hspace {1.5cm}
\includegraphics[width=.34\linewidth, angle=0]{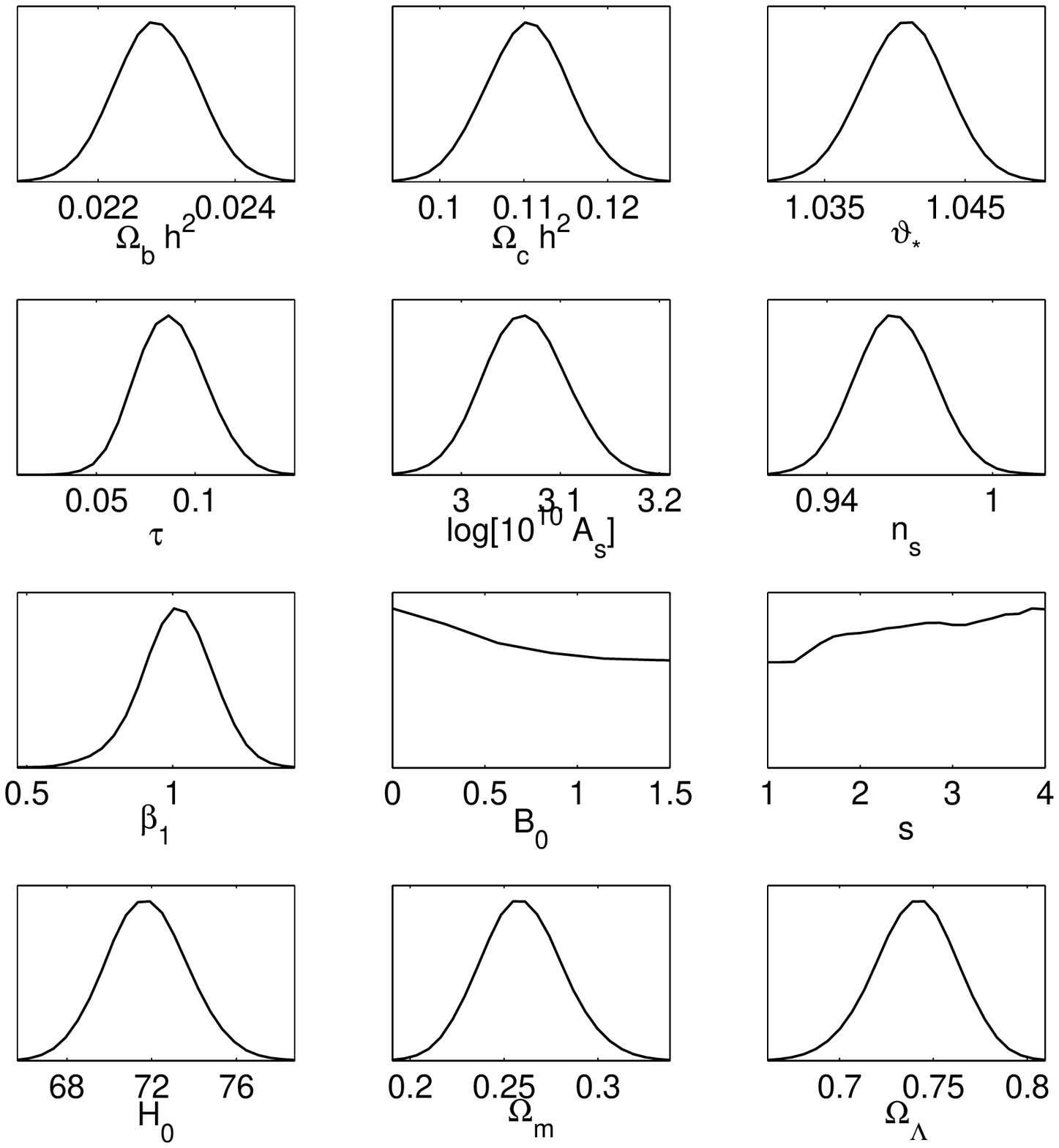}
\caption{Left: posterior likelihood distributions for the model parameters for the
  $f(R)$ case, using
  the combined CMB and ISW data (red, dashed lines) and adding also the SNe
  data (black, solid). We can see that the usual
\LCDM parameters peak around the concordance values, while the extra parameter $B_0$
  has an upper limit. The SNe tighten the constraints by reducing the
  degeneracy between $\Omega_m$ and $B_0$. Right: the same for the
  Yukawa--type case. Here we only show the full CMB+ISW+SN result. The upper
  limit on $B_0$ disappears in this case due to the removal of the corrective factor from Eq.~(\ref{par_mu}) and the additional degeneracies,
  and we can not constrain $s$ either, but a value of the coupling $\beta_1$ close to unity is required to fit the data.
}
\label{fig:posterior1}
\end{figure*}

\begin {figure}[htb]
\begin{flushleft}
\hspace{0.4cm}
\includegraphics[width=.4\linewidth, angle=0]{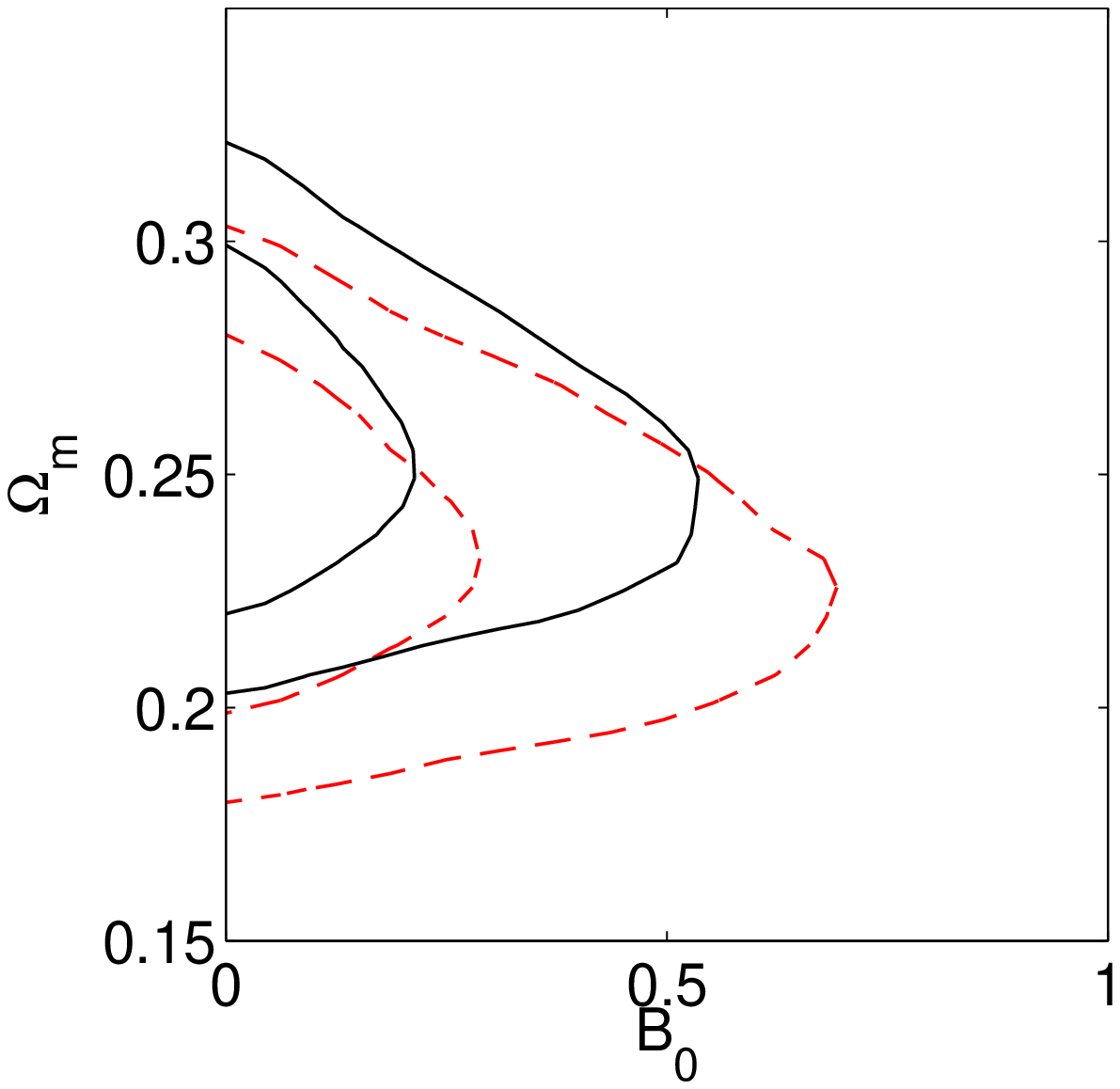}
\end{flushleft}
\includegraphics[width=1.075\linewidth, angle=0]{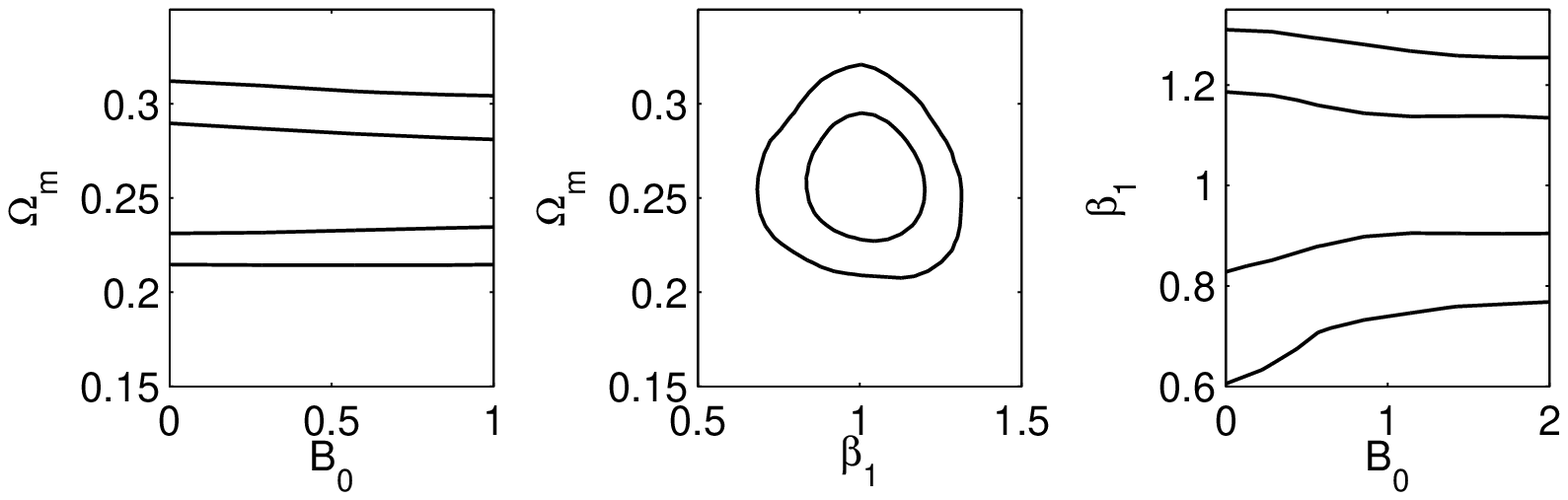}
\caption{2D posterior likelihood distributions; the 68\% and 95\% probability contours are shown. In the top panel we can see the parameters $\Omega_m, B_0$ for the
  $f(R)$ case, using
  the combined CMB and ISW data (red, dashed lines) and adding also the SNe
  data (black, solid). The SNe tighten the constraint by reducing the
  degeneracy between $\Omega_m$ and $B_0$. 
In the bottom panels, we show the Yukawa--type case for the parameters $\Omega_m,
  B_0$ (left),  $ \Omega_m, \beta_1$   (centre), and $ \beta_1, B_0$  (right).
}
\label{fig:posterior2}
\end{figure}

\section {Results \& constraints} \label {sec:constraints}

\subsection {$f(R)$ theories}

In order to be conservative, we first perform an MCMC analysis using the constraints from CMB+ISW only, as it is in general good practice to add data sets gradually, to control that there are no strong tensions between them, which would produce artificially tight posteriors.
We show in the left panel of Fig. \ref{fig:posterior1} (red dashed lines)  our constraints from this run. Here we
see the one-dimensional marginalised likelihood distributions of the
cosmological parameters.
 It can be seen that the usual \LCDM parameters have
likelihoods peaked around their standard values, while the extra parameter
$B_0$ has an upper limit, so that we find
\be
 B_0 < 0.5 \:\:\:\:\mr{or}\:\:\:\:\: \lambda_1 < 2000 \, \mr{Mpc}/h \:\: @  \, 95\%  \, \mr {c.l.}
\ee
We
have checked the convergence of this and the following results from the $R-1$ statistic between the chains, requiring always
$R-1 < 0.02$ and over 3000 estimated independent samples.
This result, which has been made possible by a full combined CMB-ISW analysis data,
 improves the previous
constraints of $B_0 < 1$ by~\cite{Song:2007da}, obtained only by considering models
with a positive ISW-matter correlation signal.

We then perform a new analysis by adding the constraints from the Union
supernovae catalogue by~\cite{Kowalski:2008ez}. By adding these constraints,
the parameter space allowed from the background expansion history becomes
narrower, and thus so become the constraints on the modification of gravity,
mainly due to the degeneracy of the theory wavelength $\lambda_1$ with other
parameters such as $\Omega_m$, as can be seen in the 2D contour plot of Fig.~\ref{fig:posterior2} (top panel). We can see the one-dimensional marginalised likelihood curves for the models in the left panel of Fig.~\ref{fig:posterior1} (black solid lines).
 The resulting constraint from the combination of
CMB+ISW+SN is

\be
 B_0 < 0.4 \:\:\:\:\mr{or}\:\:\:\:\: \lambda_1 < 1900 \, \mr{Mpc}/h \:\: @  \, 95\%  \, \mr {c.l.}
\ee

We have also tested how the results can improve by adding the latest additional priors on the value of the Hubble constant $H_0$ \cite{Riess:2009pu}; however, since the background is already well constrained by the intersection of CMB and SNe data, the improvement is marginal (at the percent level in the value of the upper limit).

\subsection {Yukawa--type dark matter interaction}
Finally we perform an analysis with three parameters to investigate theories in which dark matter particles interact via a Yukawa--type force. In this case, we directly include all data sets, i.e. CMB ISW and SNe, since we expect the marginalised constraints to be weaker due to the additional parameters.
We can see the result of the 1D posterior likelihood in the right panel of Fig.~\ref{fig:posterior1}. In this case, the upper limit on $B_0$ (or $\lambda_1^2$) is weakened and becomes uninteresting due to additional degeneracies, while
we find a constraint on the  coupling parameter
\be
0.75 < \beta_1 < 1.25 \:\: @  \, 95\%  \, \mr {c.l.}
\ee
This limit is enhanced by the lack of strong redshift evolution of the ISW signal. The result is comparable with the previous constraints by~\cite{Bean:2008ac}, although obtained under
different assumptions.
We can not find any significant bound on the additional parameter $s$.
We show in the bottom panels of Fig.~\ref{fig:posterior2} the 2D likelihood contours for $\Omega_m, B_0$, $\Omega_m, \beta_1$ and $\beta_1, B_0$. 

From the last panel of Fig.~\ref{fig:posterior2} we can better understand why the constraint on $B_0$ is now weakened. In the $f(R)$ case, the coupling parameter is fixed to $\beta_1 = 4/3$, for which high values of $B_0$ are disfavoured. When the coupling is set free in the Yukawa-type theories, its preferred values are around $\beta_1 \simeq 1$, for which we can see that $B_0$ is unconstrained.

\section {On the parametrisations and the Bayesian priors} \label{sec:priors}

In Bayesian theory, the posterior probability of a model $M$ with $N$ paramters $\mathbf{\Theta}$ given the data $ \mathbf{D}$ is obtained as 
\be
\mathcal {P} (\mathbf{\Theta}) = \frac{\mathcal {L}(\mathbf{\Theta})  \Pi (\mathbf{\Theta})}  {\mathcal {Z} (M)},
\ee
where $ \mathcal {L}(\mathbf{\Theta})$ is the likelihood function, $ \Pi(\mathbf{\Theta})$ the prior and $\mathcal {Z} (M)$ is the Bayesian evidence for the model, which we can here consider constant.
The marginalised posterior of a single parameter $\vartheta$ is then obtained integrating over the other parameters:
\be \label{eq:marginal}
\mathcal {P} (\vartheta) \propto \int \mathcal {L}(\mathbf{\Theta})  \Pi (\mathbf{\Theta}) d^{N-1} \mathbf{\Theta}.
\ee

We normally assume the priors to be flat over the sampled interval, so that they can be ignored in the marginalisation process. In this case, the posterior is simply mapped by the likelihood,  upon which assumption the MCMC technique is based.
However, if we change one parameter as $\vartheta \rightarrow \zeta (\vartheta) $, we have that a prior which was flat in $\vartheta$ will not in general be flat in $\zeta$. In other words, changing parametrisation modifies the prior from $\Pi (\vartheta)$ to
\be \label{eq:prior}
\Pi (\zeta) = \Pi (\vartheta) \frac{d \vartheta}{d \zeta},
\ee
which corresponds to a different weighting of the likelihood in Eq.~(\ref{eq:marginal}); in higher dimensions, the derivative is replaced by the Jacobian of the transformation, as described for example in \cite{Valiviita:2009bp}. If we now run an MCMC chain using the new parameter $\zeta$ as a primary parameter, we will in practice force the prior $\Pi (\zeta)$ to be flat, meaning that the prior on the old parameter  $\Pi (\vartheta)$ will be tilted by a factor $d \zeta / d \vartheta $.
 If a prior is tilted, it will cause a  downweighting of the parameter region in one direction, thus affecting the results.
All this is well known, and it has been discussed by~\cite{Lewis:2002ah}.

Clearly, if the theory is well constrained by the data, then the interval of integration will be small, and the prior will be approximately flat in any parametrisation, yielding consistent posteriors.
However, this does not apply to our case: since the current data are not yet strongly constraining MG, we have found that the choice of priors does indeed have strong effects. In this case,  we read in \cite{Lewis:2002ah} that
\begin{quote}
"...for the results of the parameter estimation to be meaningful
it is essential that the priors on the base set are well
justified..."
\end{quote}
This leaves us with the problem of deciding which of the many possible MG parameters ($B_0, \lambda_1, \lambda_1^2, \mr{Log} \lambda_1^2, ... $) is most \emph {meaningful} or \emph{well justified} to be assumed having flat priors. 
The conclusion seems uncertain since any modification of gravity is currently based on speculation without experimental backing, and it is thus hard to decide which parametrisation ought to be preferred on physical grounds. Since there is no evidence for a modification of GR, we have then decided to make the most conservative choice by presenting the parametrisation which yields the weakest constraints on this modification.

\begin{table*}[ht]
\begin{tabular}{|c|c|c|c|c|c|}
\hline
\textbf {parameter} $\bm {\zeta}$  &  $\bm {d \zeta / d \lambda_1} $ &  $\bm {d^2 \zeta / d (\lambda_1)^2} $   & \textbf{prior on $\bm {\lambda_1} $}  & \textbf {expected posterior on $\bm {\lambda_1} $}  &  \textbf {MCMC estimated} $\bm {\lambda_1} $  \textbf {@ 95\% c.l.}   \\
\hline
$\lambda_1 $            &       1          &   $0$    &   flat & baseline         &   $\lambda_1 < 1400 $ \;\, Mpc$/h$    \\
$\lambda_1^2 $          &  $\propto \lambda_1 $ &  $>0$   & positive tilt & biased to higher $\lambda_1$  &  $\lambda_1 <    1900 $ \;\, Mpc$/h$   \\
$ B_0 $                 &  $\propto \lambda_1 $ &  $>0$   & positive tilt & biased to higher $\lambda_1$  &  $\lambda_1 < 1900 $ \;\, Mpc$/h$  \\
Log $ \lambda_1^2 $     &  $\propto 1 / \lambda_1 $ &  $<0$   & negative tilt & biased to lower $\lambda_1$ & $\lambda_1 < 700 $ \;\, Mpc$/h$  \\
\hline
\end{tabular}
\caption{Effect of the priors on the estimation of the baseline parameter $\lambda_1$. Using flat priors on different parameters $\zeta$ defined by transformations whose second derivative is positive (negative) will produce effective tilted priors on $\lambda_1$. This will overestimate regions with higher (lower) $\lambda_1$, thus leading to biased posteriors. The results of MCMC chains using different parameters agree with this prediction. }
\label{tab:priors}
\end{table*}

Nevertheless, at least in the simplest $f(R)$ case with one extra parameter, we can forecast the effect of the different options by using Eq.~(\ref{eq:prior}). If we take the parameter $\lambda_1$ as our baseline choice, we can predict that by switching to flat priors in the other parameters we will have the effects summarised in Table~\ref{tab:priors}: any transformation to a parameter $\zeta$ with $d^2 \zeta / d (\lambda_1)^2 > 0 $   will tilt the priors --- and bias the posteriors --- towards higher  values of $\lambda_1$; the opposite for $d^2 \zeta / d (\lambda_1)^2 < 0 $.

Finally we compare these predictions with the results we infer on the baseline parameter $\lambda_1$ by running MCMC chains using different primary parameters; we find a good agreement with the prediction, and in particular we confirm that our choice of $B_0$ (or equivalently $\lambda_1^2$) is the most conservative one. 
If we decide that the wavelength $\lambda_1$ should be considered the most physically justified quantity, we then obtain the following stricter bounds in the $f(R)$ case using CMB+SN+ISW:
\be
\lambda_1 < 1400 \, \mr{Mpc}/h \:\:\:\: \mr{or} \:\:\: B_0 < 0.2 \:\: @ \, 95\% \, \mr {c.l.}
\ee

\section {Conclusions} \label {sec:concl}

In this paper we have presented an analysis of parametrised modified
gravity theories in light of recent data from the CMB and the large scale
structure of the Universe.

We have focused first on the $f(R)$ class of modified gravities,
which are described by a single extra parameter related to the mass of the
scalaron, which can be parametrised as its present wavelength $B_0$.
We have performed a full MCMC analysis and found some new constraints
on this parameter: in the $f(R)$ case, we have obtained $B_0 < 0.4 $ or $\lambda_1 < 1900 \, {\rm Mpc}/h $ at the 95\% c.l. when
using combined data from the ISW, the CMB and SNe.
This puts strong bounds on the parameter space of such a theory in light of
current data, therefore further reducing the possibilities for significant infrared
modifications of general relativity.

We have highlighted how  the  choice of the parametrisation, through its effect on the Bayesian priors, can be a very sensitive issue in the estimation of non-standard cosmological parameters. We have shown that by using a marginally less conservative prior we can reduce the posterior constraint to
$\lambda_1 < 1400 \, \mr{Mpc}/h $ or $  B_0 < 0.2 $ at the $ 95\% \, \mr {c.l.}$

We have then studied theories with a Yukawa--type interaction between dark matter particles; these models require two further parameters: the coupling $\beta_1$ and the time evolution of the scalaron mass $s$.
In this case we have shown how the constraints on the theory lengthscale weakens, but we obtain a bound on the coupling  $0.75 < \beta_1 < 1.25 $  at the $95\%$  c.l. when using all the data sets.

Our result proves once more that tests of structure formation and attempts to
reconstruct the evolution of the gravitational potentials are crucial to
 distinguish between modifications of gravity and dark energy theories.
Future improvements along these lines will be likely achieved by combining all
the
different available probes, such as the ISW, weak lensing, galaxy cluster
counts and peculiar velocities (see for instance~\cite{Zhao:2008bn,Zhang:2008ba,Acquaviva:2008qp,Song:2009zz,Zhao:2009fn} for forecasts) 
\footnote{During the preparation of this paper, a new  constraint on $f(R)$ theories
came from~\cite{Schmidt:2009am}, based on local clusters abundance that improves previous
constraints in~\cite{Song:2007da} by two orders of magnitude. The new constraint presented here is compatible
with this constraint, involves only linear perturbation theory and it is therefore 
fully independent and complementary. }.
\\
\\

\subsection*{Acknowledgements}
We thank Gongbo Zhao for making the MGCAMB code publicly available and for useful clarifications.
We thank the Bavarian Academy of Science and the Leibniz Computer Centre for
computational resources.
We acknowledge the Galileo Galilei  Institute for Theoretical Physics in
Florence, where this work was initiated.
TG would like to thank the CoPS group at Stockholm University for hospitality.

\bibliography{ms}

\label{lastpage}
\end{document}